\documentclass[prl,twocolumn,showpacs,floatfix]{revtex4}
\usepackage{graphicx,graphics,color,epsfig}
\usepackage{bm}
\usepackage{amsmath}
\usepackage{amssymb}

\begin{document}

\title{Cooling  Nanomechanical Resonator by Periodic Coupling to Cooper Pair Box}
\author{ P. Zhang, Y.D. Wang and C.P. Sun $^{a,b}$ }
\affiliation{Institute of Theoretical Physics, Chinese Academy of Sciences, Beijing,
100080, China}

\begin{abstract}
We propose and study an active cooling mechanism for the nanomechanical
resonator (NAMR) based on periodical coupling to a Cooper pair box (CPB),
which is implemented by a designed series of magnetic flux pluses threading
through the CPB. When the initial phonon number of the NAMR is not too
large, this cooling protocol is efficient in decreasing the phonon number by
two to three orders of magnitude. Our proposal is theoretically universal in
cooling various boson systems of single mode. It can be specifically
generalized to prepare the nonclassical state of the NAMR.
\end{abstract}

\maketitle

\textit{Introduction---} Recently nanomechanical resonators (NAMR) have been
fabricated with high quality factors from ($10^{2}$ to $10^{5}$) and large
fundamental frequencies (in the range of MHz $\sim $GHz) \cite%
{apl,nature1,namr-prl}. The NAMR has been shown as a good candidate for
exploring various mesoscopic quantum phenomena at the boundary between
classical and quantum realms. Until now NAMRs have been used in generating
entangled states \cite{entanglement}, demonstrating quantum non-demolition
measurement \cite{prb} and progressive quantum decoherence \cite{wang}, and
implementing a two qubit quantum gate \cite{gate}.

The quantum nature of NAMR has been exhibited by the accurate measurement
near the standard quantum limit \cite{nature2,science}. But in most cases,
to fully utilize the quantum wealths provided by NAMR it is necessary to
cool the NAMR to its ground state. There have been some schemes for cooling
NAMR \cite{hopkins,zoller2,zoller1}. Some of them are based on coupling with
Josephson junction (JJ) qubit and make use of feedback control and sideband
cooling techniques. Characterized by the maximal ratio between the average
number of phonons before and after cooling, the highest efficiency of some
schemes \cite{zoller2,zoller1} can be achieved when the initial number of
phonons $N_{\text{th}}$ of the NAMR is large enough ($N_{\text{th}}\gtrsim
10^{-1}$). The cooling effect is evidently decreased as $N_{\text{th}}$ gets
very small.

Motivated by the existing investigations mentioned above, we suggest a
straightforward mechanism to cool the NAMR. Our scheme is also based on the
coupling with the Cooper pair box (CPB), which is considered as a
controllable two-level system. Different from the existing schemes, our
cooling protocol works efficiently in small $N_{\text{th}}$ regime ($N_{%
\text{th}}\lesssim 10^{1}$). In our scenario, the interaction takes place
periodically between the CPB and the NAMR. Before the interaction takes
place in each cycle, the CPB is always set to its ground state so that it
can absorb some energy from the NAMR during interaction period. A similar
method has been used by to cool the microwave cavity \cite{haroche,nature3}.
In principle, the present proposal can be generalized for cooling any single
mode boson system via the coupling with a two-level system.

One can intuitively compare our cooling mechanism with a classical analog.
To cool a thermal box one can put a piece of ice into it and then drain the
melt down water. It will take away part of the heat in the box. Naively, one
can freeze the drained water into ice outside the box in some way and then
place it back into the box. Repeat this process again and again until the
box reaches the desired temperature. In our scheme, the CPB prepared in the
ground (excited) state can be imagined as the ice (water) and the NAMR as
the thermal box in the classical analog. However, due to quantum coherence,
the mechanism of our cooling protocol is not as naive as this
\textquotedblleft ice-and-box" analog, because the energy loss of box is
irreversible due to the second law of thermodynamics. The substantial
difference of our protocol from the above classical analogue is the coherent
oscillation of energy exchange between the qubit and the bosonic mode. In a
sense, our scheme is more related to a type of quantum heat engine \cite%
{scully1}.
\begin{figure}[h]
\begin{center}
\includegraphics[width=3.5cm,height=4cm]{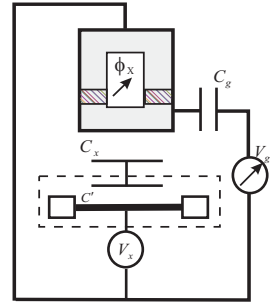} \vspace{0.3cm}
\end{center}
\caption{ The coupling system of the NAMR (within the rectangle of dashed
lines) and the CPB: the bang-bang coupling is implemented by a predesigned
periodic series of magnetic flux $\Phi _{x}$ threading through the CPB. }
\end{figure}

\textit{Model for our cooling protocol\textbf{\ }--- }In our NAMR-CPB
composite system shown schematically in Fig.1, the NAMR is directly
connected to a CPB consisting of two Josephson junctions. The external
magnetic flux $\Phi _{x}$ threads through the SQUID, which can be used to
adjust its effective Josephson energy. $C_{J}$ is the Josephson capacitance,
$V_{g}$ the tunable gate voltage, $C_{g}$ the gate capacitance, $V_{x}$ the
bias voltage on the resonator, and $C^{\prime }$ the effective capacitance
of the NAMR. $C_{x}$ denotes the distribution capacitance between the CPB
and the NAMR. Let $E_{c}=e^{2}/(2\left( C_{x}+C_{g}+C_{J}\right) $ be the
charge energy unit and \ $n_{g}\left( x\right) =\left( C^{\prime }\left(
x\right) V_{x}+C_{g}V_{g}\right) /2e$ \ be the total gate charge.

The dependence of charging energy $4E_{c}\left( n_{c}-n_{g}\left( x\right)
\right) ^{2}$ on $x$ results in the coupling between the NAMR with free
Hamiltonian $p^{2}/(2m)+m\omega _{0}^{2}x^{2}/2$ and the CPB with
controllable Josephson tunnelling energy $-2E_{J}\cos \left( \pi \Phi
_{x}/\Phi _{0}\right) \cos \theta .$ Here, $n_{c}$ denotes the number of the
excess Cooper pair on the island while its conjugate variable is the phase
difference $\theta $ of the two sides of each junction. Usually the
fluctuation of $x$ is much smaller than the distance $d$ between the NAMR
and the CPB. At the exact resonance point defined by $\left(
C_{x}V_{x}+C_{g}V_{g}\right) /2e=N/2$\ where $N$ is odd, the CPB acts as a
two level system, a qubit. We denote two linearly independent charge states
by $\left\vert 1\right\rangle =\left\vert n_{c}=(N-1)/2\right\rangle $ and $%
\left\vert 2\right\rangle =\left\vert n_{c}=(N-1)/2\right\rangle $. Under
the two level approximation and the rotation wave approximation (RWA), we
write the above Hamiltonian as the Jaynes-Cummings (JC) form \cite{zoller1}%
\begin{equation}
H=E_{J}\cos \left( \pi \frac{\Phi _{x}}{\Phi _{0}}\right) \sigma _{z}+\omega
_{0}a^{\dag }a+g\left( a\sigma _{+}+h.c\right) .  \label{aaa}
\end{equation}%
Here, $\sigma _{z}\left( \sigma _{+},\sigma _{-}\right) $ are defined with
respect to the new basis $\{\left\vert e\right\rangle =\left( \left\vert
1\right\rangle +\left\vert 2\right\rangle \right) /\sqrt{2},\left\vert
g\right\rangle =\left( \left\vert 1\right\rangle -\left\vert 2\right\rangle
\right) /\sqrt{2}\}$; $a\left( a^{\dag }\right) $ are the phononic creation
and annihilation operators of the NAMR mode with the effective coupling
constant $g=4E_{c}n_{x}\sqrt{1/2m\omega _{0}}/d$ where $%
n_{x}=C_{x}V_{x}/(2e) $.

In our protocol, the above JC type interaction is assumed to take place
periodically with rate $r_{a}$. Then it is switched off after the duty cycle
interval $\tau $ of the order of $g^{-1}$. This on-and-off switching can be
realized by the magnetic flux $\Phi _{x}$. In fact, at the exact resonance
point, the JJ tunnelling energy $E_{J}\cos \left( \pi \Phi _{x}/\Phi
_{0}\right) $ is just the energy level spacing. It will be switched to the
value resonant with the NAMR during the duty cycle $\tau $ and far-off
resonant outside this period. The similar manipulation has been used to
create the non-classical photon state based on the superconductor devices
\cite{liu}

To implement our proposal we have two main tasks to realize the periodical
coupling: (a). Switching on and off the interaction between the CPB and the
NAMR periodically, and (b). Preparing the CPB to its ground state before the
interaction takes place in each cycle. Both tasks can be accomplished via
tuning the magnetic flux $\Phi _{x}$\ with time. In fact, the gate charge
fluctuation induced relaxation rate $\Gamma \left( \omega \right) =\pi
\alpha _{g}\omega \left[ \coth \left( \omega /2k_{B}T\right) +1\right] /2$
of the CPB at temperature $T$\ \cite{zoller1} can be well controlled by
varying the flux $\Phi _{x}$ since $\omega =E_{J}\cos \left( \pi \Phi
_{x}/\Phi _{0}\right) $. Here, $k_{B}$\ is the Boltzmann constant, $\alpha
_{g}$\ is about $2e^{2}R\left( C_{x}^{2}+C_{g}^{2}\right) /\left( \pi
(C_{x}+C_{g}+C_{J})^{2}\right) $\ , and $R$\ the fluctuation impedance of $%
V_{g}$\ and $V_{x}$. Outside the duty cycle $\tau $, one can switch the
energy spacing $\omega $ of CPB to a large value to satisfy $\omega \pm
\omega _{0}>>g$. In this case, the CPB-NAMR interaction is effectively
switched off because of the far-off resonance (this can be deduced without
RWA) while the decay process is enhanced. Thus the CPB can be prepared well
in its ground state $\left\vert g\right\rangle $\ for the up-coming
interaction period.

\textit{Master equation approach in steady states---} We assume
the coupling strength $g$\ is much stronger than $\Gamma \left(
\omega _{a}\right) $\ and $\kappa $\ where $\kappa $\ is the decay
rate of NAMR, and the interaction period $\tau $ is so short that
$\Gamma \left( \omega _{a}\right) \tau <<1$, $\kappa \tau <<1.$ In
this case\textbf{\ }both the decay of CPB during the duty cycle
and the NAMR-environment coupling\textbf{\ }can be omitted.
Therefore, if the interaction is switched on at instance $t_{l}$,
the reduced density operator $\rho \left( t_{l}+\tau \right) $ of
the NAMR after a time interval $\tau $ can be obtained through the
action of the superoperator $M\left( \tau \right) $ on the reduced
density operator $\rho \left( t_{l}\right) $ at instance $t_{l}$,
i.e. $\rho \left( t_{l}+\tau \right) =M\left( \tau
\right) \left[ \rho \left( t_{l}\right) \right] $, which is defined as $%
M\left( \tau \right) \left[ \rho \left( t_{l}\right) \right] =$Tr$_{\text{a}}%
\left[ \exp (-i\hat{h}\tau )\rho \left( t_{l}\right) \otimes \left\vert
g\right\rangle \left\langle g\right\vert \exp (i\hat{h}\tau )\right] $Tr$_{%
\text{a}}$ denotes tracing over the variables of CPB. $\hat{h}=ga\sigma
_{+}+h.c$ is the JC type Hamiltonian (\ref{aaa}) at resonance in the
interaction picture. Without any dissipation, the exact solution of the
resonant JC model gives the explicit recursions
\begin{equation}
p_{n}\left( t_{l}+\tau \right) =\left\vert c_{g,n}\left( \tau \right)
\right\vert ^{2}p_{n}\left( t_{l}\right) +\left\vert c_{e,n}\left( \tau
\right) \right\vert ^{2}p_{n+1}\left( t_{l}\right)
\end{equation}%
for the diagonal elements $p_{n}=\left\langle n\right\vert \rho \left\vert
n\right\rangle $ of $\rho \left( t_{l}+\tau \right) $ for the number $%
\left\vert n\right\rangle $ state of NAMR phonon. Here, $c_{e,n}\left( \tau
\right) =\sin \left( g\tau \sqrt{n+1}\right) $ and $c_{g,n}\left( \tau
\right) =\cos \left( g\tau \sqrt{n}\right) $ come from the exact solution of
the resonant JC model.

With the presence of the dissipation of NAMR, the evolution of $\rho $ can
be depicted by the course gained master equation
\begin{equation}
\frac{d\rho }{dt}=r_{a}\left[ M\left( \tau \right) -1\right] \rho +L\left[
\rho \right] .  \label{aa}
\end{equation}%
The super operator $L$ in the above equation attributes to the dissipation
and is defined as $L\left[ \rho \right] =-(\kappa /2)N_{\text{th}}\left(
aa^{\dagger }\rho -2a^{\dagger }\rho a+\rho aa^{\dagger }\right) -(\kappa
/2)\left( N_{\text{th}}+1\right) \left( a^{\dagger }a\rho -2a\rho a^{\dagger
}+\rho a^{\dagger }a\right) $ where $N_{\text{th}}=\left[ \exp \left( \omega
_{0}/k_{B}T\right) -1\right] ^{-1}$ is the average number of phonons of NAMR
at temperature $T$ before cooling. The master equation (\ref{aa}) was
initially presented for the case that the interaction between the two level
system and the single mode oscillator is "turned on" randomly \cite%
{scully,orszag}. In the case where the interaction in our scheme is
periodically "turned on", this equation can also lead to a correct stable
solution.

Without detailed computations, the average number of phonons $\left\langle
n\right\rangle _{s}=$Tr$\left[ \rho _{s}a^{\dagger }a\right] $ of NAMR in
the steady state $\rho _{s}$ can be obtained from the above master equation:
\begin{equation}
\left\langle n\right\rangle _{s}=N_{\text{th}}-\left( r_{a}/\kappa \right)
\Delta n.  \label{d}
\end{equation}%
where $\Delta n=$Tr$\left\{ a^{\dagger }a\left[ 1-M\left( \tau \right) %
\right] \rho _{s}\right\} $. Since Tr$\left\{ a^{\dagger }a\left[ 1-M\left(
\tau \right) \right] \rho _{s}\right\} >0$ can be proved with the definition
of $M(\tau )$, we can conclude that $\left\langle n\right\rangle _{s}<N_{%
\text{th}}$, i.e., the NAMR can always be cooled when the steady state is
reached.

\textit{Dynamic process of cooling and the fluctuation of number of
phonons--- }The steady state solution of Eq. (\ref{aa}) gives the phonon
population
\begin{equation}
p_{n}^{s}=p_{0}^{s}\prod_{l=1}^{n}\frac{N_{\text{th}}l}{(N_{\text{th}%
}+1)l+\left\vert c_{e,l-1}\right\vert ^{2}r_{a}/\kappa }.  \label{bb}
\end{equation}%
Here, $p_{n}^{s}$ is $n$-th diagonal element of $\rho _{s}$ and $p_{0}^{s}$
is determined by the normalization condition $\sum_{i=0}^{\infty
}p_{i}^{s}=1 $ \cite{scully}. By a virtue of a numerical computations, we
can describe exactly the evolution of vacuum state probability $p_{0}$ and
average number $\left\langle n\right\rangle $ of phonons in detail. Fig.2
demonstrates these results with experimentally rational parameters $N_{\text{%
th}}\approx 1.7$, $r_{a}/\kappa =133$ and $g\tau =\pi /8$. It is shown that,
at the time $t_{a}$ satisfying $r_{a}t_{a}\sim 60$, the steady solution (\ref%
{bb}) is reached. In this sense the distribution of the number of phonons in
the data does not vary significantly after time $t_{a}$.

\begin{figure}[h]
\begin{center}
\includegraphics[width=6cm,height=4.5cm]{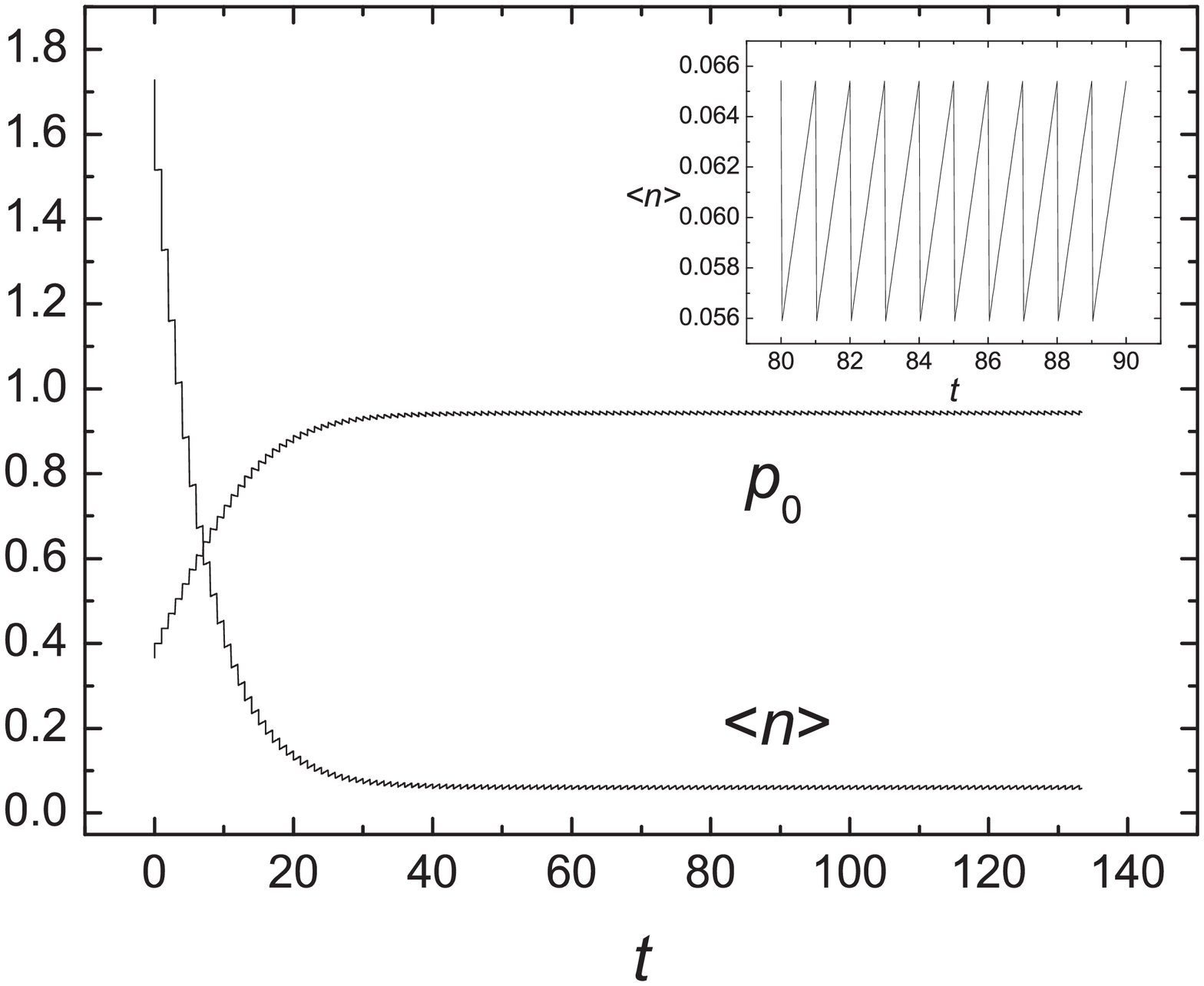}
\end{center}
\caption{ The time evolution of the average phonon number $\langle n\rangle $
and the vacuum state probability $p_{0}$. The unit of time is $1/r_{a}$. The
fluctuation of phonon number in the steady state is shown in the inset.}
\end{figure}

Apparently, we can improve the cooling effect by increasing $%
r_{a}$. For a given value of $r_{a}/\kappa $, our cooling scheme works well
when the average number of phonons $N_{\text{th}}$ before cooling is small
enough, i.e., $N_{\text{th}}<<r_{a}/\kappa $. Under this condition, we can
achieve the maximum cooling effect by setting the duty cycle $\tau =\pi
/\left( 2g\right) $ so that $\left\vert c_{e,0}\right\vert =1$. In this case
the average number of phonons after cooling is%
\begin{equation}
\left\langle n\right\rangle _{s}\approx p_{1}^{s}\approx N_{\text{th}}\kappa
/r_{a}.  \label{ee}
\end{equation}%
This results implies that the number of phonons is reduced by a factor of $%
\left( r_{a}/\kappa \right) $. However, if $N_{\text{th}}$ is comparable
with (or larger than) $\left\vert c_{e,0}\right\vert ^{2}r_{a}/\kappa $, our
scheme does not work well. For $N_{\text{th}}+1>>r_{a}/\kappa $, we have $%
p_{n}\approx p_{0}\left[ N_{\text{th}}/(N_{\text{th}}+1)\right] ^{n}$, which
implies that the average number of phonons $\left\langle n\right\rangle _{s}$
in the steady state is very close to $N_{\text{th}}$, the number of phonons
before cooling. The average number $\left\langle n\right\rangle _{s}$ of
phonons after cooling is drawn against $N_{\text{th}}$ in Fig. 3. It shows
that, with $r_{a}/\kappa =10^{2}$ and $\left\vert c_{e,0}\right\vert =1$, we
have $\left\langle n\right\rangle _{s}\approx 10^{-2}N_{\text{th}}$ \ for $%
N_{\text{th}}\lesssim 1$; with $r_{a}/\kappa =10^{3}$ and $\left\vert
c_{e,0}\right\vert =1$, we have $\left\langle n\right\rangle _{s}\approx
10^{-3}N_{\text{th}}$ for $N_{\text{th}}\lesssim 10$ and $\left\langle
n\right\rangle _{s}\approx 10^{-2}N_{\text{th}}$ \ for $N_{\text{th}%
}\lesssim 10^{2}$ respectively.

\begin{figure}[h]
\begin{center}
\includegraphics[width=7cm,height=5.5cm]{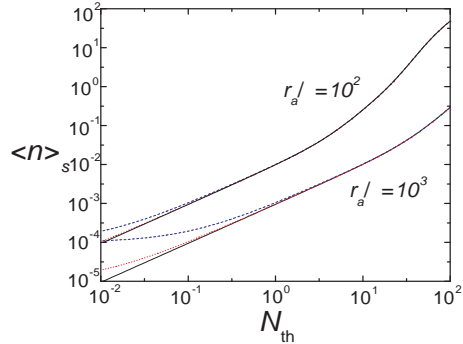}
\end{center}
\caption{ (color online) The cooling effect diagram. Here we assume $\protect%
\alpha _{g}\approx 1\times 10^{-4}$. The finial phonon number $\langle
n\rangle _{s}$ is drawn as a function of the initial phonon number $N_{\text{%
th}}$ when $r_{a}/\protect\kappa =10^{2}$ or $10^{3}$. In the solid (black)
line, the thermal excitation of the CPB is not considered. In the dashed
(blue) line, we assume the thermal excitation probability of the CPB is $%
p=10^{-4}.$ In the dotted (red) line, we assume $p=10^{-5}.$ }
\end{figure}

It is also noted that, since the interaction between the CPB and the NAMR is
switched on and off again and again, and the master equation (\ref{aa}) is
obtained via a coarse granulation approach, the number of phonons will have
a fluctuation $\Delta n=$Tr$_{\text{N}}\left\{ a^{\dagger }a\left[ 1-M\left(
\tau \right) \right] \rho _{s}\right\} $ even in the steady state. In the
inset of Fig. 2, the fluctuation $\Delta n$ is shown in the evolution curve
of the number of phonons in the steady state.

\textit{Experimentally feasible predications---} We consider a NAMR with
frequency $\omega _{0}$ =$2\pi \times 10^{2}$MHz ($0.5\mu $eV) and quality
factor $Q=2\times 10^{5}$. Our cooling protocol can be realized with the
following experimentally accessible parameters, the decay rate $\kappa $ = $%
\pi \times 10^{-3}$MHz, the Josephson energy $E_{J}\approx $ $4\pi \times
10^{4}$MHz ($100\mu $eV), the Coulomb charging energy $E_{C}\approx 320\mu $%
eV (i.e. $C_{\Sigma }\approx 250$aF) ,the mutual capacitance $%
C_{x}=C_{g}\approx 20$aF, the impedances $R\approx 50\Omega $ and the gate
voltage $V_{x}\approx 0.25$V. With these parameters, the interaction
strength $g$ is estimated to be $2\pi \times 10$MHz, the number of Cooper
pair $n_{x}$ is about 15 and $\alpha _{g}$ $\approx 1\times 10^{-4}$.

When the magnetic flux $\Phi _{x}$ in the Hamiltonian (\ref{aaa}) is tuned
to about $0.498\Phi _{0}$ on resonance, the interaction takes place. We set
the interaction duration $\tau $ = $2.5\times 10^{-8}$s so that $\left\vert
c_{e,0}\left( \tau \right) \right\vert =1$. If the temperature $T$ is $0.01$
Kelvin, during the duty cycle, the decay rate $\Gamma \left( \omega
_{0}\right) $ of CPB is about $0.56$MHz. Just after the interaction, the
magnetic flux should be turned off to maximize the energy spacing. Then the
CPB will decay to its ground state by a rather large decay rate $\Gamma
(E_{J})\approx 40$ MHz. The CPB then decays quickly in $\tau ^{\prime
}=0.275\mu $ s. This elementary procedure is repeated at a frequency $%
r_{a}\approx 3$ MHz. Namely, the interaction takes place every $%
T_{1}=1/r_{a}=0.3\mu $s. The time is sufficiently long for completing the
interaction and the relaxation in each elementary procedure. Therefore,
omitting the dissipation effect of the CPB, we can estimate from Eq. (\ref%
{ee}) that the number of phonons after cooling has the magnitude of $%
\left\langle n\right\rangle _{s}\sim N_{\text{th}}/\left( r_{a}/\kappa
\right) \sim 10^{-3}N_{\text{th}}$.

In the above discussion, the parameter of crucial importance $\kappa $ is
determined by the quality factor $Q$ and the resonance frequency $\omega
_{0} $ of the NAMR. As illustrated above, for $Q\sim 2\times 10^{4}$, we end
up with $r_{a}/\kappa =10^{2}$, and the cooling effect would be much worse.
A careful analysis of the cooling cycle reveals that the cooling effect is
also influenced by the dissipation of the CPB during the duty cycle. For
high initial temperatures, we can replace the factor $\left\vert
c_{e,l-1}\right\vert ^{2}$ in formula (\ref{bb}) by $\left\vert
c_{e,l-1}\right\vert ^{2}F_{l-1}$ where $F_{l-1}$ is the fidelity defined as
$F_{l-1}=1-\Gamma \left( \omega _{0}\right) \int_{0}^{\tau }\left\vert
c_{e,l-1}\left( t^{\prime }\right) \right\vert ^{2}dt^{\prime }$ from a
first order perturbation calculation. This leads to a more precise phonon
distribution $p_{n}$ in the steady state. For parameters corresponding to $%
\alpha _{g}\approx 1\times 10^{-4}$ as given above, the correction to $p_{n}$
due to the above perturbation effect of the inclusion of CPB dissipation can
be omitted when $N_{\text{th}}\lesssim 10^{2}$. On the other hand, the
coupling of the CPB to the environment may cause the CPB to jump from the
ground state to the excited state with a rate $\Gamma \left( \omega
_{0}\right) $, and subsequently emit a phonon via coherent Rabi oscillation.
In the low temperature case, this probability can be estimated as $N_{\text{%
th}}\Gamma \left( \omega _{0}\right) \int_{0}^{\tau }\left\vert
C_{e,0}\left( \tau -t^{\prime }\right) \right\vert ^{2}dt^{\prime }$\ $=N_{%
\text{th}}\allowbreak \Gamma \left( \omega _{0}\right) \tau /2$. In our
example as mentioned above, this value is about $7\times 10^{-3}N_{\text{th}%
} $. The detailed study on the influence of the dissipation effect of the
CPB on our cooling protocol will be given in our future work.

More generally, for the case of finite temperatures, there is always a
thermal excitation for the CPB. In fact, even when off duty, the CPB decays
to a mixed state with probability $p=1/\left( 1+\exp [E_{J}/k_{B}T]\right) $
in the excited state $\left\vert e\right\rangle $ and $\left( 1-p\right) $
in the ground state $\left\vert g\right\rangle $. For a finite value of $p$,
the elements of the steady state density matrix $\rho _{s}$ should be
modified to be
\begin{equation}
p_{n}^{s}=p_{0}^{s}\prod_{l=1}^{n}\frac{N_{\text{th}}l+p\left\vert
c_{e,l-1}\right\vert ^{2}r_{a}/\kappa }{(N_{\text{th}}+1)l+\left( 1-p\right)
\left\vert c_{e,l-1}\right\vert ^{2}r_{a}/\kappa }.
\end{equation}%
this leads to a reduction of the the cooling efficiency. In Fig.
3, the influence of this thermal excitation probability is
illustrated. In the low temperature limit, the average number of
phonons emitted by the CPB via this mechanism during the duty
cycle is still $p$. Combined with the analysis of the previous
paragraphs, the lower limit phonon number fluctuation at steady
state due to thermal excitations and dissipations of the CPB can
be estimated as $1/\left( 1+\exp [E_{J}/k_{B}T]\right)
+N_{\text{th}}\Gamma \left( \omega _{0}\right) \tau /2$, which
vanishes in the limit of large $E_{J}$ and low $\Gamma \left(
\omega _{0}\right) $. This can also be considered as the lower
limit of the average phonon number after cooling.

\textit{Conclusion with remarks about relations to maser --- }We should
notice that the periodic cooling of the NAMR can be understood as an
\textquotedblleft inverse" of the \textquotedblleft maser" mechanism. In the
usual maser process, the input excited atoms (molecules) can coherently heat
the cavity and then \textit{coherently accumulates} photons in a single
state to enhance cavity field in a quantum way. In the present protocol, the
NAMR is cooled by the CPB in its ground state. It should be emphasized that,
with the same setup and operations similar to the above protocol, a "NAMR
maser" can be devised if the CPB is initially prepared in its excited state $%
\left\vert e\right\rangle $ before the interaction is switched on. Namely,
we can prepare the NAMR in the non-classical state with the number of
phonons in super-Poissonian or sub-Poissonian distribution. On other hand,
in many protocols of two-qubit quantum logic gates based on different
physical systems, the bosonic mode in its ground state can serve as a
quantum data bus to transfer quantum information from its coupled qubit to
another, or to entangle two qubit at a distance. Therefore, the protocol in
this paper may have potential applications in quantum information theory.

The authors also thank the useful discussions with Y.X. Liu and
L.You. This work is supported by the NSFC with grant Nos.
90203018, 10474104 and 60433050. It is also funded by the National
Fundamental Research Program of China with Nos. 2001CB309310 and
2005CB724508.


\begin{thebibliography}{a}
\bibitem[a]{email} Electronic address:suncp@itp.ac.cn

\bibitem[b]{www} Internet www site: http:// www.itp.ac.cn/\symbol{126}suncp

\bibitem{apl} A.N. Clelandand M.L. Roukes, Appl. Phys.Lett. \textbf{69},
2653 (1996).

\bibitem{nature1} X.M.H. Huang, et al.,Nature(London) \textbf{421}, 496
(2003).

\bibitem{namr-prl} A. Gaidarzhy, et al, Phys. Rev. Lett. \textbf{94}, 030402
(2005)

\bibitem{entanglement} A. D. Armour, et al., Phys. Rev. Lett. \textbf{88},
148301 (2002);

\bibitem{prb} E. K. Irish and K.Schwab, Phys. Rev. \textbf{B68}, 155311
(2003);

\bibitem{wang} Y.D. Wang, Y.B.Gao, C.P. Sun, Euro. Phys. J. B \textbf{40} ,
321 (2004).

\bibitem{gate} X. Zouand W. Mathis, Phys. Lett. A \textbf{324}, 484 (2004).

\bibitem{nature2} R.G.Knobel and A.N. Cleland, Nature (London) \textbf{424},
291 (2003).

\bibitem{science} M.D. LaHaye, et al., Science \textbf{304}, 74 (2004).

\bibitem{hopkins} A. Hopkins, et al., Phys. Rev. B \textbf{68}, 235328
(2003).

\bibitem{zoller2} I. Wilson-Rae, et al., Phys. Rev. Lett. \textbf{92},
075507 (2004).

\bibitem{zoller1} I. Martin, et al., Phys. Rev. B \textbf{69}, 125339 (2004).

\bibitem{haroche} A.Rauschebeutel, et al., Phys.Rev.Lett.\textbf{83}, 5166
(1999).

\bibitem{nature3} S. Nogues, et al., Nature (London), \textbf{400}, 239
(1999).

\bibitem{scully1} M.O. Scully, et al., Science \textbf{299}, 862 (2003).

\bibitem{liu} Yu-xi Liu, L. F. Wei, and Franco Nori, Europhys. Lett. \textbf{%
67} , 941(2004).

\bibitem{scully} M.O. Scully and M.S. Zubiary, \textit{QuantumOptics}
(Cambridge University Press, Cambridge, U.K., 1997).

\bibitem{orszag} M. Orszag, \textit{Quantum optics : including noise
reduction, trapped ions, quantum trajectories, and decoherence} (Springer,
Berlin, 2000).
\end{thebibliography}
\end{document}